\documentstyle[aps,prb,multicol,epsf]{revtex} 
\begin{document} 
\widetext
\title{\bf Ground state properties and excitation spectra of 
non-Galilean invariant interacting Bose systems}

\author{G.~Jackeli\cite{byline} and J. ~Ranninger}

\address{Centre de
Recherches sur les Tr\`es Basses Temp\'eratures, 
Laboratoire
Associ\'e \'a l'Universit\'e Joseph Fourier, 
\\ Centre National de la
Recherche Scientifique, BP 166, 38042, Grenoble 
C\'edex 9, France}

\date{\today} 
\maketitle 
\draft 
\begin{abstract}
We study  the  ground state properties and the excitation spectrum of 
bosons which, in addition to a short-range repulsive two body potential, 
interact through the exchange of some dispersionless bosonic modes.
The latter induces a time dependent (retarded) boson-boson interaction
which is attractive in the static limit. 
Moreover the coupling with 
dispersionless modes introduces a reference frame for the moving boson 
system and hence breaks the Galilean invariance of this system.
The ground state of such a system is depleted {\it linearly}
 in the boson density
due to the zero point fluctuations driven by the retarded part 
of the interaction.
Both quasiparticle (microscopic) and compressional (macroscopic) sound
velocities of the system are studied.  
The microscopic sound velocity is calculated up the second order in 
the effective two body interaction in a perturbative treatment,
 similar to that of  Beliaev for 
the dilute weakly interacting Bose gas.
The hydrodynamic equations are used to obtain the macroscopic 
sound velocity. We  show that  these  velocities 
are identical within  our perturbative approach.
We present  analytical results for them  
in terms of two dimensional parameters -- an effective interaction strength
and an adiabaticity parameter --
which characterize the system. 
We find that due the presence of several competing effects, which
 determine the speed of the sound of the system, three 
qualitatively different regimes can be in principle realized 
in the parameter space and discuss them on  physical grounds.  
   
\end{abstract}


\begin{multicols}{2}

\narrowtext

The superfluid properties of Bose-systems have been studied and greatly 
understood since the early fifties.\cite{BEC-TRENTO} While an ideal  
Bose gas condenses but does not exhibit superfluidity,  an interacting 
Bose gas with repulsive interaction does. 

The microscopic description of excitation spectrum of weakly interacting 
dilute Bose gas (WIDBG)  has been first considered by Bogoliubov.~\cite{bog}
It   has been shown that the interaction among the particles
modifies the character of low  lying elementary excitation in the condensed 
state of the system~\cite{bog} and as a result long-wavelength excitations  
obey the linear dispersion law, thus having a sound-wave-like character
rather then particle-like as in the normal state of the system.
Within this lowest order treatment, referred to
nowadays as Bogoliubov approximation,
the microscopic sound velocity was found to be identical to the  
macroscopic, compressional, sound velocity of the system. 
Going beyond this lowest order theory, Beliaev has developed a 
perturbation expansion of the WIDBG \cite{bel,pop,grif} and considered 
the second order corrections to the Bogoliubov results. 
His results confirmed the equivalence of microscopic and macroscopic 
sound velocities.\cite{bel,hp} 
This equivalence has been generalized to all orders of perturbation theory 
by Gavoret and Nozi\`eres in Ref.~\onlinecite{gn} invoking explicitly the
Galilean invariance of the system.

In a recent paper \cite{jr} we have 
addressed ourselves to a class of systems composed of 
two coupled bosonic subsystems: propagating  bosons,
interacting through a short-range repulsive interaction, and 
local  dispersionless bosonic degree's of freedom 
(for  brevity hereafter termed as 
$\it phonons$ ) to which they are coupled.
The  coupling to the latter not only supplements the
short range two body potential of bosons by a time dependent counterpart
(attractive at low energies) but also  introduces a reference frame for 
the moving boson system and hence breaks the Galilean  
invariance of this system. 
The  aim of the present work is to study the ground state properties and
the excitation spectrum of such a system,
focusing  on the renormalization of long-wavelength sound-like
excitation
due to the presence of the competing interactions in the system
mentioned above.

We first determine 
the microscopic sound velocity from the single-particle excitation spectrum.
The latter has been calculated in Ref.\onlinecite{jr} within the   
Beliaev-Popov \cite{bel,pop} second-order perturbation theory 
generalized to our system. 
Here we consider the continuum version of the model 
discussed in  Ref.\onlinecite{jr} and 
show that the leading order terms in a density expansion
can be treated analytically. 
We discuss the modification of the perturbation expansion 
due to the presence of retarded interaction 
and the range of validity of our treatment. To calculate the macroscopic sound velocity we base ourself on the hydrodynamic equations, 
explicitly incorporating the
non-Galilean invariance of the system. We find  
that both velocities are identical  up to the order 
considered in  the present approximation.

The Hamiltonian which describes the above discussed scenario is given by
\begin{eqnarray}
H &=& \sum_{\bf q} (\varepsilon_{\bf q}-\mu)b^{\dagger}_{\bf q} b_{\bf q} 
+\frac{g}{2N} \sum_{\bf k, k', p} b^{\dagger}_{\bf k} b^{\dagger}_{\bf k'} 
b_{\bf k'-p}b_{\bf k+p} \nonumber \\
&+&  \omega_0 \sum_{\bf q}
(a^{\dagger}_{\bf q}a_{\bf q}+\frac{1}{2})
-\frac{\alpha  \omega_0}{\sqrt N} \sum_{\bf k,q} 
b^{\dagger}_{\bf k} b_{\bf k+q}[a^{\dagger}_{\bf q} + a_{\bf -q}]~,
\label{H}
\end{eqnarray} 
 where $b^{({\dagger})}_{\bf k}$  and $a^{({\dagger})}_{\bf q}$ denote 
the boson and phonon annihilation (creation) operators, 
$\varepsilon_{\bf q}=q^2/(2m)$, $m$ and $\mu$ is the boson mass and chemical
potential, respectively.
The interaction among bosons is assumed to be properly renormalized
and is given by   $g=4\pi a/m$. We note, that working with 
a renormalized interaction, one has to be careful to avoid double counting of
the diagrams which have already been included in the ladder sum for the renormalized interaction, that defines the $s$-wave scattering length $a$ of two particles
in vacuum. The strength of the boson-phonon coupling is given  
by $\alpha  \omega_0$, ($\alpha$  being  dimensionless parameter)
and  $\omega_0$ denotes the frequency of the dispersionless phonon mode.

In our  treatment we assume both, the dilute limit of the boson system and 
weak boson-phonon coupling in order to set up a perturbation theory for 
the above introduced Hamiltonian at $T=0$. 
The phonon mode can be effectively 
integrated out leading to the effective two body potential
$\Gamma_{{\bf q},\omega}=g+\alpha^2\omega_{0}^{2}{\cal D}^{(0)}
_{{\bf q} ,\omega}$, where ${\cal D}^{(0)}_{{\bf q}, \omega}=2\omega_{0}/(\omega^2-\omega_{0}^{2})$ is the bare phonon Green's function.
The phonon meditated  interaction vanishes at high 
frequencies as $1/\omega^{2}$ and hence no special renormalization or
ultraviolet cut-off is required  in order 
to perform calculations to a given order.

The lowest order approach to the above Hamiltonian is constructed within the
Bogoliubov scheme \cite{bog}: by separating out in Eq.~(\ref{H}) 
the condensate part of  the Bose field 
$b_{\bf k} =\hat b_{\bf k}+n_{c}^{\frac{1}{2}} \delta_{k,0}$ 
($n_{c}$ being the condensate fraction)
and keeping only the bilinear part of the resulting Hamiltonian.
To this order of approximation 
one obtains two branches of the excitation spectrum of the system \cite{jr}:
\begin{eqnarray}
\omega^2_{\pm,{\bf q}} &=& {E^2_{{\bf q}} + \omega^2_0 \over 2} 
\pm \frac{1}{2} \sqrt {[E^2_{{\bf q}} -  \omega^2_0]^2 + 
16 \alpha^2  \omega_0^3 \varepsilon_{{\bf q}}n_{c}}
\label{spectr}
\end{eqnarray} 
with 
$E_{{\bf q}}=[\varepsilon_{\bf q}^{2}+2gn_{c}\varepsilon_{\bf q}]^\frac{1}{2}$
denoting the spectrum of Bogoliubov quasi-particles in the absence of 
the coupling with phonons. 

When the bosons condense, the phonons,  being initially coupled to 
the boson density (symmetry restoring variable in the case of gauge 
symmetry breaking), get hybridized with the Goldstone mode.
The resulting two normal modes (\ref{spectr}) describe  Bogoliubov type  
excitations and Einstein phonons. For  momenta close to where the level 
crossing of bare excitation spectra occurs neither mode is predominantly 
a phonon nor a Bogoliubov quasi-particle. 
In the long-wavelength limit the lower branch  reduces to a sound-like 
dispersion
$\omega_{-,{\bf q}}\simeq v_{0}q$, where 
$v_{0}=[(g-2\alpha^2\omega_{0})n_{c}/m]^{\frac{1}{2}}$ 
is a characteristic sound velocity
with the repulsive interaction having being  reduced by the static part of the attractive phonon mediated 
interaction.

The Bogoliubov scheme amounts to taking into account  
collisions between  condensate 
quasi-particles,  between condensate and out-of-condensate quasi-particles, 
while neglecting scattering among out-of-condensate quasi-particles.
The interaction among the particles depletes the ground state condensate
leading to a finite density $\tilde{n}$ of non condensed particles.
For the weakly interacting dilute Bose gas, the depletion is given by 
$\tilde{n}=n-n_{c}\sim n^{3/2}$ and is small in the low density limit
$n\ll 1$. Beliaev\cite{bel,grif} 
 has considered the second ordered corrections to the 
Bogoliubov result and has shown that the expansion parameter of the theory is the so called gas parameter $an^{1/3}$.
The latter measures the effective range of the interaction  $a$ 
(the $s$-wave scattering length) relative to the inter-particle distance $d\sim n^{1/3}$. All the second-order corrections to the Bogoliubov results 
(for example corrections to the chemical potential $\mu$, squared sound velocity $v^{2}$) appear  with higher powers in density ($\sim n^{3/2}$) and 
are small in the dilute limit even if the interaction is strong.
As it follows, these arguments no longer apply to the case we consider.
The reason is the following: even in the vanishing density limit the system 
does not reduce to the non interacting one and one ends up with the well known polaron problem. In  other words, due to the retarded character 
of phonon mediated  interaction, one particle can interact with itself, by creating the lattice deformation at given time and absorbing it later on.

A generalization of the Beliaev-Popov \cite{bel,pop,grif} second-order perturbation theory to the presently studied  problem  has been discussed by us in Ref.\onlinecite{jr}.
As we have already mentioned, here we restrict ourselves to the leading 
order in density contributions,  which allows us to obtain analytical results. 
We calculate the corrections to the 
chemical potential, microscopic sound velocity and discuss the expansion parameter and  the limit of the validity of the perturbation theory.

\begin{figure}
\epsfysize=44mm
\centerline{\epsffile{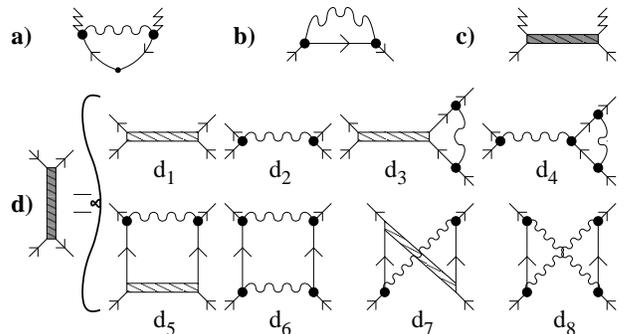}}
\caption{a) Leading order in density contribution to  ground state depletion,
b) normal component of the boson self-energy due to the coupling with phonons,
c) anomalous component of the boson self-energy, 
d) renormalized  four point vertex.}
\label{f1}
\end{figure}

To begin with we calculate the depletion of the ground state $\tilde{n}$.
It is given by the number of out-of-condensed particles and vanishes for 
the ideal Bose gas. In Fig.~1a  we express graphically the linear  in density 
contribution to the depletion.  The  straight and wave lines stand for the 
boson and phonon bare Green's functions, respectively,  
the dot represents the boson-phonon vertex and the  zig-zag line 
the condensate (given by a factor $n_{c}^{1/2}$).
This contribution is exclusively due to the phonon mediated retarded 
interaction.
It vanishes for the instantaneous interaction 
since in this case the end points (see Fig.~1a) would correspond to the same time, 
which involves both   ``particle'' and ``hole'' type excitations. 
Only the former one is non vanishing for $T=0$. 

The corresponding  analytical expression of the 
diagram shown in Fig.~1a is as follows 
\begin{eqnarray}
\tilde{n}&=&i\alpha^2\omega_{0}^{2}n_{c}
\int\frac{d\omega}{2\pi}\frac{d{\bf q}}{(2\pi)^3}
\left[{\cal G}^{(0)}_{{\bf q},\omega}\right]^2{\cal D}_{-{\bf q},-\omega}^{(0)}
\nonumber\\
&=&\alpha^2\omega_{0}^{2} n_{c}\int\limits_{0}^{\infty}\frac{q^2d q}{2\pi^2}\frac{1}
{(\varepsilon_{q}+\omega_{0})^2}=\frac{\alpha^2(2m\omega_{0})^{\frac{3}{2}}n_{c}}{8\pi}.   
\label{depl1}
\end{eqnarray}
In Eq.~(\ref{depl1}) 
${\cal G}^{(0)}_{{\bf q},\omega}=(\omega-\varepsilon_{q}+i\eta)^{-1}$ denotes the bear boson Green's function. As a result, to the leading order in density, the ground state depletion  and the condensate fraction takes the form:
\begin{equation}
\tilde{n}=\frac{\gamma\lambda}{2} n_{c}+O\left(n_{c}^{\frac{3}{2}}\right)~,\;\;
n_{c}\simeq\frac{n}{1+\gamma\lambda/2}~,
\label{depl}
\end{equation}
where we have introduced two dimensionless parameters: 
$\lambda=2\alpha^2\omega_{0}/g$ which is the
strength of phonon mediated interaction relative to the direct boson-boson  
repulsion and $\gamma=a/a_{0}$ being the ratio of two characteristic lengths --
the $s$-wave scattering length $a$ and $a_{0}=1/q_{0}$ with 
$q_{0}=\sqrt{2m\omega_{0}}$ being the momentum at which level crossing of
the boson and the phonon bear spectrum occurs.
We note that the parameter $\gamma$ always appear in the combination
$\gamma\lambda$ which does not explicitly involve the boson-boson repulsion
and is equivalent to the so-called Migdal parameter in the electron-phonon problem. However, as it follows, the interpretation of our  results in terms of 
$\gamma$ and $\lambda$ is more transparent and we keep this notations,
rather using the conventional one.

The renormalization of the spectrum given by Eq.~(\ref{spectr}), due to the second order corrections can be obtained from the poles of the dressed boson propagator.
The diagonal (${\cal G}_{{\bf q},\omega}$)  and off-diagonal 
($\hat{\cal G}_{{\bf q},\omega}$) elements  of the boson Green's function 
are  expressed  through the normal ($\hat\Sigma_{{\bf q},\omega}$)
 and anomalous ($\Sigma_{{\bf q},\omega}$) components of the self-energy 
\begin{eqnarray}
{\cal G}_{{\bf q},\omega}&=&[\omega+\epsilon_{\bf
q}+\Sigma_{{\bf q},-\omega}]/{\text D}_{{\bf q},{\omega}},\;\;
\hat{ \cal G}_{{\bf q},\omega}=[-\hat\Sigma_{{\bf
q},\omega}]/{\text D}_{{\bf q},{\omega}},\nonumber\\
{\text D}_{{\bf q},{\omega}}&=&[\omega-{\cal A}_{{\bf q},\omega}]^2-
[\epsilon_{\bf q}+S_{{\bf
q},\omega}][\epsilon_{\bf q}+S_{{\bf q},\omega}+2\hat\Sigma_{{\bf
q},\omega}],
\label{DB}
\end{eqnarray}
where $\epsilon_{\bf q}=\varepsilon_{\bf
q}-\mu$,
${\cal S}_{{\bf q},\omega}=[\Sigma_{{\bf q},\omega}+\Sigma_{{\bf q},-\omega}-2\hat\Sigma_{{\bf q},\omega}]/2$ and 
${\cal A}_{{\bf q},\omega}=[\Sigma_{{\bf q},\omega}-\Sigma_{{\bf
q},-\omega}]/2$ are  symmetric  
and antisymmetric functions  in $\omega$,  respectively.
Invoking the Hugenholtz-Pines theorem \cite{hp} we have the relation 
$\mu=\Sigma_{0,0}-{\hat\Sigma}_{0,0}=S_{0,0}$.

Expanding the self-energies in the long-wavelength, low
frequency limit as follows
\begin{eqnarray}
{\cal A}_{{\bf q},\omega}={\bar{\cal A}}\omega,\;\;\;\;
{\cal S}_{{\bf q},\omega}-\mu={\bar{\cal S}}_{1}\omega^{2}+{\bar{\cal S}}_{2}q^2
\label{expan}
\end{eqnarray}
we obtain  the microscopic sound velocity from Eq.~(\ref{DB}) as
\begin{eqnarray}
v^2= \frac{\hat\Sigma_{0,0}(1+{\bar{\cal S}}_{2})}
{m[(1+|{\bar{\cal A}}|)^2+2|{\bar{\cal S}}_{1}|\hat\Sigma_{0,0}]}.
\label{DBmis}
\end{eqnarray}

Let us first discuss the self-energy expansion parameters defined 
in Eq.~(\ref{expan}). As  we have already mentioned,
 we are interested in the leading order (linear in density) contributions.
Since the anomalous self-energy $\Sigma_{0,0}$ entering into Eq.~(\ref{DBmis})
vanishes   linearly with density, there will be no contribution
to this order coming  from  ${\bar{\cal S}}_{1}$ and  
only those from the remaining $|{\bar{\cal A}}|$ and ${\bar{\cal S}}_{2}$
(which has nonzero value in the limit of vanishing density) will contribute
to this order. 
The first non-zero contribution to ${\bar {\cal A}}$  comes 
from the diagram presented in Fig.~1b. 
Substituting the bare boson and phonon Green's functions in the corresponding 
formula of this  diagrams, and expanding the result of the 
integration in powers of the external frequency $\omega$,
after straightforward algebra
one obtains ${\bar{\cal A}}=\gamma\lambda/2$. 
One also  verifies, that  for the dispersionless Einstein phonons 
considered here,  the  phonon mediated self-energy (see Fig.~1b)
being local in space is  momentum independent. 
Hence,  for the momentum expansion parameter
 ${\bar{\cal S}}_{2}$ (\ref{expan}) one has ${\bar{\cal S}}_{2}=0$ 
within the same order of approximation.

At this point, it is interesting to check whether the exact relations\cite{gn1}
\begin{equation}
|{\bar{\cal A}}|=
\partial{\tilde n}/{\partial n_{c}}~,\;\;\;  
{\bar{\cal S}}_{2}q^2=n_{c}^{-1}[e_{q}^{\prime}-e_{0}^{\prime}]
\label{exrel}
\end{equation}
derived in Ref.\onlinecite{gn} still holds, given the  above obtained perturbative
results.  In Eq.~(\ref{exrel})  $e_{0}^{\prime}$ and  $e_{q}^{\prime}$ stand for  the energy 
per unit volume of the system at rest and the system  moving uniformly with a speed
$v=q/m$, respectively (prime denotes that the contribution from the condensate phase is eliminated).
With the help of Eq.~(\ref{depl}), 
one can easily verify that the first relation Eq.~(\ref{exrel})
is recovered and not violated by
the perturbative theory. As for the second relation, invoking the Galilean invariance of the system it has been linked in Ref.\onlinecite{gn} 
to the change in kinetic energy of the moving system
leading to ${\bar{\cal S}}_{2}=n_{c}^{-1}{\tilde n}/(2m)$. For the present case [see Eq.~(\ref{depl})] this gives a 
finite value ${\bar{\cal S}}_{2}=\gamma\lambda/(4m)$
instead of zero as has been discussed above. 
However, as it will be shown  in details below, 
due to the non-Galilean invariance in the present case there is an additional contribution to the energy of the  moving system,
 stemming from the  retarded boson phonon interaction. 
Considering both these contributions, 
it turns out that to within this order of approximation of our perturbative treatment
 they cancel each other and one recovers 
${\bar{\cal S}}_{2}=0$.

To the leading order in a density expansion the anomalous 
self-energy is given by the diagram shown in Fig.~1c, corresponding to
$\hat\Sigma_{0,0}={\tilde\Gamma}n_{c}$ with  
${\tilde\Gamma}$ denoting a four point vertex  dressed 
by the phonon mediated interaction (see Fig.~1d)  with all external
momenta and frequencies taken to be zero.
The renormalization of the four point vertex in the second order is 
shown in Fig.~1d. The diagrams $d_3$ and $d_4$ 
from  Fig.~1d
represent the 
renormalization of the boson-boson and the  boson-phonon vertices, 
respectively.
The diagrams $d_5$ and $d_6$   describe  the two-body $t$-matrix
renormalization arising from the
phonon-mediated interaction. The last two diagrams ($d_7$ and $d_8$)
 represent the 
exchange processes and, like the diagrams  for vertex renormalization,    
are nonzero due to the time dependence of the phonon mediated interaction.
We note that there are no equivalent diagrams 
with both interaction lines being the boson-boson renormalized interaction
$g$, since either these contributions 
involve backward propagators of the bosons in the vacuum and hence
vanish or they have already been included in $g$ and do not 
have to be double counted.
 After substituting the bare boson and phonon Green's functions in the corresponding 
standard analytical formulas of the above discussed diagrams, 
all the integration can be done analytically 
and one arrives to the following expression for 
$\tilde\Gamma$ (see Ref.\onlinecite{gamma}):
\begin{equation}
\tilde\Gamma=g[1-\lambda+4\gamma\lambda-17\gamma\lambda^2/8] 
\label{gamma0}
\end{equation}
Based on  the above obtained result the sound velocity from Eq.~(\ref{DBmis}) 
can be written as:
\begin{eqnarray}
v^2\simeq\frac{gn}{m}\left[1+\frac{\lambda}{8}(-8+20\gamma-5\gamma\lambda)
+O(\gamma^2\lambda^2)\right].
\label{mis}
\end{eqnarray}  
This expression  implies the perturbation expansion in powers of 
$\lambda$ and  $\gamma\lambda$. The 
dilute limit does not necessary imply the convergence of the 
perturbation expansion  
and we  have to assume moreover $\lambda<<1$ and $\gamma<<1/\lambda$.

Before providing a   physical interpretation of these   results, let us 
 examine the macroscopic sound velocity of the system to within  
the same order of perturbation theory. 
For that purpose we start with the well known
equations of motion for the  two conjugate dynamical variables: the  number operator
$n$ and the macroscopic phase of the condensate $\phi$. It reads as:
\begin{equation}
\frac{\partial n}{\partial t}=\frac{\partial e_{n,\phi}}{\partial\phi},\;\;\;
\frac{\partial \phi}{\partial t}=-\frac{\partial e_{n,\phi}}{\partial n},
\label{eom}
\end{equation}
where $e_{n,\phi}$ denotes  the energy of the system per unit volume.
The next step is to consider a
hydrodynamic potential flow of the boson system with the superfluid velocity
defined as ${\bf v}_{s}={\bf\nabla} \phi/m$. 
Then the change in energy of the system at finite but small  $v_{s}$
can be written, without loss of generality,  as $\delta e_{n,v_{s}}=e_{n,v_{s}}-e_{n,0}=\Lambda mv_{s}^{2}/2$, where $\Lambda$  is the phase stiffness 
of the superfluid system.
Taking the {\it grad} from both sides of  the second equation (\ref{eom}) 
and using the thermodynamic relation
$\partial e_{n,\phi}/\partial n=\mu$ one arrives to 
the Josephson relation for  the superfluid velocity field
$m\partial{\bf v}_{s}/{\partial t}=-{\bf\nabla}\mu$.
The  two equations of motion (\ref{eom}) can be combined into:
\begin{equation}
\frac{\partial^2 n}{\partial t^2}=c^2{\bf \nabla}^2 n,
\label{sound}
\end{equation}
with $c^2=\Lambda/\chi$ being the speed of macroscopic sound, and $1/\chi=\partial \mu/\partial n$  
the compressibility of the system.

 In the case of Galilean invariant system
the change of the system's energy is exclusively
 due to the kinetic energy term in the
 Hamiltonian and is given by  $\delta e_{n,v_{s}}=\delta e_{n,v_{s}}^{\rm kin}=nmv_{s}^2/2$. 
The potential energy, being only a function of the relative positions of the particles, 
does not change when the boson system  moves with finite velocity.
Then the phase stiffness is given by $\Lambda=n/m$ and one recovers the well known expression 
for the speed of compressional sound $c^2=n/(m\chi)$.
However, in the case of coupling with phonons, the system is no longer Galilean invariant 
since then this coupling introduces the reference frame for the 
moving system of bosons.  In other words, the phonon mediated interaction, being retarded, 
depends on the frequency and hence on the velocity of the particles.
Therefore, there will be an additional contribution to $\delta e_{n,v_{s}}$
due to the boson-phonon coupling  $\delta e_{n,v_{s}}^{\rm B-P}$.
\begin{figure}
\epsfysize=24mm
\centerline{\epsffile{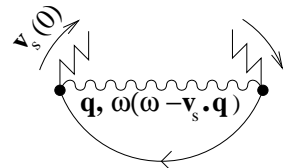}}
\caption{Ground state energy correction due to the boson-phonon coupling.
The quantities in the  parentheses  correspond to the frame  moving together 
with the boson system.}
\label{f2}
\end{figure}

In Fig.~2 we present graphical expression for the first nonzero contribution to 
$\delta e_{n,v_{s}}^{\rm B-P}$. In the frame at which the phonons are at rest, 
the condensate mode has a 
finite momentum equal to ${\bf v}_{s}/m$. To calculate $\delta e_{n,v_{s}}^{\rm B-P}$
 it is convenient to perform a Galilean transformation to the frame in which the boson 
system is at rest. This  amounts to replacing 
 the frequency $\omega$ appearing in the  phonon propagator in the diagram shown in Fig.~2 by
$(\omega-{\bf v}_{s}{\bf q})$. Evaluating this contribution and expanding 
the result for small ${v}_{s}$ one finds 
$\delta e_{n,v_{s}}^{\rm B-P}=-\gamma\lambda nmv_{s}^{2}/[2(2+\gamma\lambda)]$. 
Combining the two contributions,  $\delta e_{n,v_{s}}^{\rm kin}$ and 
$\delta e_{n,v_{s}}^{\rm B-P}$, we arrive at $\delta e_{n,v_{s}}=\Lambda^{*} m^{2}v_{s}^{2}/2$, where
the renormalized phase stiffness is given by  
\begin{equation}
\Lambda^{*}=n/[(m(1+\gamma\lambda/2)].
\label{lambda}
\end{equation}
The next step is calculate compressibility of the system. To this end we 
evaluate the chemical potential by using the Hugenholtz-Pines relation
$\mu=\Sigma_{0,0}-{\hat\Sigma}_{0,0}$.\cite{hp} Examining the 
various diagrams  for $\Sigma_{0,0}$ and  $\hat{\Sigma}_{0,0}$
one  finds that  most of them cancel pairwise, and to leading order
(linear in density)  we have 
$\mu=const+(g-2\alpha^2\omega_{0})n+\mu^{(2)}$. The constant term describes the negative density independent shift of the chemical potential due to the coupling with phonons and does not contribute to the compressibility.
The second term is the first order contribution given by the reduced 
boson repulsion arising from  the static attractive  part of phonon 
mediated interaction.
The last term describes the second order corrections and can be written as 
$\mu^{(2)}=[2d_5+d_6+2d_7+d_8]n_{c}$ (see Ref.\onlinecite{mu}) where  
$d_i$ denotes the diagrams shown in Fig.~1d and the corresponding values are given in Ref.\onlinecite{gamma}. 
The final result for compressibility is written
as 
\begin{equation}
\chi^{-1}=1-\lambda-3\gamma\lambda -9\gamma\lambda^2/8.
\label{com}
\end{equation}
 
Finally,  upon using Eqs.(\ref{sound}), (\ref{lambda}), and (\ref{com})
 one obtains the macroscopic sound velocity:
\begin{eqnarray}
c^2\simeq\frac{gn}{m}\left[1+\frac{\lambda}{8}(-8+20\gamma-5\gamma\lambda)
+O(\gamma^2\lambda^2)\right],
\label{mas}
\end{eqnarray}  
which is  thus identical to the  microscopic one (\ref{mis}).

Let us now discuss our main findings. 
Our perturbative expansion assumes $\lambda \ll 1$ and $\gamma\lambda \ll 1$. The
first condition is necessary to insure the stability of the condensate against the collapse when the attractive interaction overcompensates the repulsive one,
the second is equivalent to the one of the Migdal approximation in 
the electron-phonon problem. For small $\lambda \ll 1$ one can neglect the
last, higher order, term in Eq.~(\ref{mas}) and obtains the following expression
for the sound velocity  
$c^2\simeq c_{0}^{2}[1+\lambda(-1+5\gamma/2)]$, with $c_{0}^{2}=gn/m$ being the 
speed  the sound in the absence of any boson-phonon coupling.
Analyzing this expression as a function of $\gamma$ one finds three different
regimes of behavior:\\ 
i) for $\gamma<2/5$, corresponding to a small  phonon mode frequency, 
the sound velocity decreases when the boson-phonon 
coupling is switched on,\\ 
ii) for  $\gamma=2/5$ the linear in $\lambda$  
contribution is exactly zero, and the sound velocity does not depend on 
$\lambda$, provided the latter is small as we have already assumed, and\\ 
iii) if $\gamma>2/5$, corresponding to a high  phonon mode frequency,
the coefficient in front of $\lambda$  becomes positive and the sound becomes 
stiffer when the boson-phonon coupling is switched on.

While  intuitively the softening of the sound   would be naturally expected,
 the increase of the sound is a rather  surprising result.
It can be understood as follows. 
The softening of the sound is due to two effects:
 First the phonon mediated interaction, being attractive,
reduces the boson-boson repulsion, and second 
it enhances the boson mass. However, if the reduction of the 
repulsive interaction in the lowest order is not strong enough, 
it could be overcompensated by the enhancing
of the boson density vertex due to the coupling with phonons. 
Moreover, this enhancement could also overcompensate 
the reduction of sound velocity
due to the enhancement of the boson mass. Both
effects, mass enhancement and vertex renormalization have the same origin,
 being only  driven by the time dependence of the  phonon mediated 
interaction. 
Hence these effects  have the same order of magnitude and, acting in 
different directions, strongly compete. This is manifest in the
 three different regimes found in our perturbative approach. 
We would like to point out,  that all regimes can in principle be  realized within the range of 
validity of the present treatment.

In our previous paper,\cite{jr} 
analyzing numerically Beliaev-Popov self energies for different values of 
$\alpha$ (boson-phonon coupling strength), without restricting ourself
 to the leading order terms in density expansion, discussed here,
 we came to the conclusion that the sound velocity was 
practically unaffected by the coupling with the phonons.
This  situation which corresponds to the second regime discussed above.

In conclusion, we have studied the ground state properties and the excitation spectrum of the  system of 
bosons which, in addition to short-range repulsive two body potential, 
interact through the exchange of the other dispersionless bosonic modes, 
for brevity termed as phonons.
We have assumed the dilute limit of the boson system and 
weak boson-phonon coupling and considered the second order perturbation theory,
equivalent of the Beliaev-Popov treatment of the weakly interacting dilute Bose gas. 
We have shown that the ground state of such a system is depleted {\it linearly} 
in the boson density due to the zero point fluctuations driven by retarded 
part of the interaction. 
We have derived analytical expressions for the microscopic and 
macroscopic sound velocities of the system and shown that up to the order of our perturbative approach they are identical. 
We found 
 that three physically different regimes 
can be realized in the parameter space of the system and discussed them on  physical grounds.  
While one can think of a number of phenomena
which might be described in terms of the model we considered, 
we leave the application of our study to realistic systems for some 
future work.

We are indebted  to P. Nozi\`eres
for useful discussions. G. J. acknowledges support from a  
{\it Bourse Post-Doctoral du  minist\`ere de l'Education nationale, 
                     de la recherche et de la technologie} and from an 
 INTAS Program, Grant No 97-0963 and No 97-11066.

\end{multicols}

\begin{references}
\bibitem[*]{byline} On leave from Institute of Physics,
Georgian Academy of Sciences,  380077 Tbilisi, Georgia.
\bibitem{BEC-TRENTO} ``Bose-Einstein Condensation'' edt. Griffin, 
Snoke and Stringari, Cambridge University Press (1995).
\bibitem{bog} N. N. Bogoliubov, J. Phys. USSR {\bf 5}, 23 (1947).
\bibitem{bel} S. T. Beliaev, Sovjet Physics JETP {\bf 7}, 299 (1958).
\bibitem{pop} V.N. Popov,
``Functional Integrals in Quantum Field theory and Statistical Physics'', 
Ch. 6, Reidle, Dortrecht (1983).
\bibitem{grif} For a review see, 
H. Shi and A. Griffin, Phys. Rep. {\bf 304}, 1 (1998).
\bibitem{hp} N. M. Hugenholtz and D. Pines, Phys. Rev. {\bf 116}, 489 (1959).
\bibitem{gn} J. Gavoret and P. Nozi\`eres, Ann. Phys. (New York) {\bf 28},
349 (1964).
\bibitem{jr} G. Jackeli, and G. Ranninger, Phys. Rev. B {\bf 63},
184512 (2001).
\bibitem{gn1} See Eqs. (4.7) and (4.9) in Ref.\onlinecite{gn}. 
We point out, that in deriving these equations
no Galilean invariance has been invoked in Ref.\onlinecite{gn} and hence 
they   still hold  in present case.
\bibitem{gamma} First we point out, that the  diagram given in Fig.~1a 
that has been already evaluated Eq.~(\ref{depl1}) is the basic element 
of the diagrams
$d_3$, $d_4$ and $d_7$ shown in Fig.~1d. Based on this observation one finds
$d_3=d_7=g\gamma\lambda/2$ and $d_4=-2\alpha^2\omega_{0}\gamma\lambda/2$.
One farther verifies, that diagram $d_5$, which can be obtained from 
$d_7$ by reversing an arrow  of one of the boson propagators, is twice bigger.
This leads to $d_5=2d_7=g\gamma\lambda$ since in the former one ($d_5$) both 
two poles (in lower and upper half of the complex plane) of the phonon 
propagator contribute.
The same holds for $d_6$ and $d_8$ -  and it is straightforward to 
show  $d_6=2d_8=-3\alpha^2\omega_{0}\gamma\lambda/2$. 
Putting all these contributions together and noting that the diagrams 
$d_3$, $d_4$, $d_5$, and $d_7$ enters with the symmetric factor 2, one obtains
${\tilde\Gamma}=d_1+d_2+8d_3+2d_4+3d_8$. Substituting in the last expression those  
corresponding values of the diagrams, discussed above, one recovers 
Eq.~(\ref{gamma0}).
\bibitem{mu} The second order corrections to the chemical potential 
$\mu^{(2)}$ have been  discussed in details in Ref.\onlinecite{jr}. 
The diagrammatic representation of the  contribution  to $\mu^{(2)}$
due to the boson phonon coupling
 is shown in Fig.~7 of that paper. 
The contributions due to the boson-boson
interaction is obtained by replacing the phonon propagator by the boson-boson interaction line in these diagrams. 
The leading order in density terms, can then be obtained by  replacing
 the dressed boson propagators in these diagrams by corresponding perturbation series and keeping only the leading, linear in density terms. 
In this way the same
diagrams as those shown in Fig.~1d of the present paper are generated and 
one recovers the analytical formula for $\mu^{(2)}$ presented in the text.
\end{references}
\end{document}